\documentclass[reprint,twocolumn,superscriptaddress,aps]{revtex4-2}
\usepackage{mathtools}
\usepackage{verbatim}
\usepackage{amsmath}
\usepackage[font=small]{caption}
\usepackage{subcaption}
\usepackage{appendix}
\usepackage{esint}
\usepackage{geometry}
\usepackage{lipsum}
\usepackage{amsfonts}
\usepackage{booktabs}
\usepackage{braket}
\usepackage{bm}
\usepackage{textcomp}
\usepackage{hyperref}
\usepackage{amssymb}

\begin{document}

\title{Split-Post Re-entrant Microwave Displacement Transducer with Quadratic Readout}

\author{Sonali Parashar, Jeremy F. Bourhill, Maxim Goryachev and Michael E. Tobar}
\affiliation{Quantum Technologies and Dark Matter Labs, Department of Physics, University of Western Australia, 35 Stirling Highway, Crawley, WA 6009, Australia.}

\begin{abstract}
We investigate a microwave-cavity-based displacement readout employing a split-post geometry to measure the motion of a dielectric membrane. Due to symmetry, the cavity response to membrane displacement is inherently quadratic when the membrane is positioned at the centre of the posts. We characterise this behaviour by driving the membrane with a piezoelectric actuator at both central and off-centre positions and we estimate the drive-to-displacement transfer function using the independently calibrated frequency-to-voltage response of the interferometric readout. When the membrane is located at the centre of the cavity and driven, the system exhibits the largest quadratic output, measured at the second harmonic of the membrane acoustic frequency. As the membrane is moved away from the centre, the response transitions from predominantly quadratic to predominantly a linear response at the membrane acoustic frequency. Quadratic optomechanical coupling is a key requirement for displacement - squared readout and, in the quantum regime, for measurements sensitive to mechanical energy or phonon number. The present work therefore establishes the split-post geometry as a promising platform for microwave-mechanical transduction, providing a practical route toward future experiments aimed at probing quantised mechanical motion and energy-sensitive readout schemes.
\end{abstract}

\maketitle

\section{Introduction}

The interaction between mechanical elements and microwave cavities is central to the field of cavity optomechanics, offering powerful means to manipulate photons via their coupling to phonons \cite{aspelmeyer2014cavity}. Such hybrid systems operate across a wide temperature range, from room temperature down to cryogenic conditions (4 K and millikelvin), and have enabled the development of ultra-sensitive displacement transducers \cite{barzanjeh2022optomechanics,arcizet2006high,rocheleau2010preparation, whittle2021approaching, huang2024room}. These systems have application in not only in displacement sensing, but also sensing of force \cite{westphal2021measurement} and mass \cite{fuchs2024measuring}. A variety of architectures have been employed to detect mechanical motion in microwave-based systems. Common approaches use 3D lumped-element (LCR) microwave resonators incorporating a mechanical element \cite{ barzanjeh2022optomechanics}, or qubit-based dispersive readouts where the microwave resonator couples to a superconducting qubit, which acts as the readout for mechanical motion \cite{shnyrkov2023rf, yuan2015large, lescanne2020irreversible, romanenko2023search, ilinskaya2024flux}. For low-noise operation at millikelvin temperatures, junction-based parametric amplifiers have been developed to enhance signal-to-noise ratios \cite{mutus2013design, lecocq2017nonreciprocal, macklin2015near, faramarzi2024near}, to achieve the standard quantum limit, and similar techniques extend to readouts based on SQUID-based amplifiers \cite{tobar1995characterizing, whittle2021approaching}.

Mechanical resonators come in diverse forms, including bulk acoustic wave (BAW) \cite{goryachev2012extremely,galliou2013extremely,Yiwen2017,Yiwen2024,Diamandi:2025aa} and surface acoustic wave (SAW) devices \cite{satzinger2018quantum}, thin membranes \cite{carvalho2017sensitivity, locke2002properties, locke2004measurement, kumar2024optomechanically}, and even lower-dimensional photonic structures \cite{tang20172d, aly2020defect}. These can be integrated on-chip as micro-/nano-mechanical hybrid resonators coupled to microwave or optical photons \cite{kotler2017hybrid, clerk2020hybrid, marinkovic2021hybrid}. The electromagnetic cavities themselves may be realised as lumped-element re-entrant modes \cite{mcallister2017higher, carvalho2019piezo, bourhill2020generation, parashar2024upconversion, carvalho2014piezoelectric,Davis2024}, whispering gallery modes \cite{tobar2002proposal, barroso2004reentrant, schliesser2010cavity, goryachev2014controlling}, or travelling-wave structures \cite{macklin2015near, lecocq2017nonreciprocal}, including transmission line geometry \cite{faramarzi2024near}. %or magnetic-mode cavities for spin coupling \cite{goryachev2014high}.
Crucially, the opto-mechanical coupling between the microwave cavity mode and the mechanical displacement typically produces a linear coupling; however, higher-order dependencies of the microwave frequency on displacement can be engineered \cite{jayich2008dispersive, thompson2008strong, burgwal2020comparing, brandao2020entanglement, xuereb2013selectable, cattiaux2020beyond}. When cooled close to its motional ground state \cite{schneeloch2023principles, whittle2021approaching}, the mechanical mode enters the quantum regime \cite{yang2019ground}, where its transitions obey Fermi’s golden rule \cite{tobar2024detecting,tobar2025detecting}. Achieving ground-state cooling allows operation beyond the thermal noise floor \cite{saulson1990thermal, locke2004measurement}, and with sufficiently high cooperativity, a quadratic readout can effectively realise a multi-level quantum system \cite{paraiso2015position, jayich2008dispersive, brandao2020entanglement, thompson2008strong, burgwal2020comparing}. Such systems exhibit phenomena including strong opto-mechanical coupling \cite{parashar2024upconversion}, parametric amplification \cite{stanwix2005test, kumar2024optomechanically}, ground-state cooling, and squeezing \cite{marinkovic2021hybrid} and can store quantum information in entangled states \cite{jayich2008dispersive, paraiso2015position}, quantum non-demolition measurement of photon number \cite{ludwig2012enhanced, stolyarov2023photon} and different non-linear interactions \cite{saiko2024optomechanical}.

Macroscopic mechanical resonators provide powerful platforms for probing the classical–quantum boundary~\cite{bushev2019testing,parashar2024upconversion} and for sensing weak signals in the kHz–MHz frequency range. Their large participating mass and high mechanical quality factors make them well suited to applications including high-frequency gravitational-wave detection~\cite{goryachev2014gravitational,aggarwal2021challenges,goryachev2021rare,campbell2021searching,campbell2023multi}, resonant-mass gravitational antennas~\cite{van1997grail,blair1995high}, and searches for dark matter through its coupling to acoustic and electromagnetic modes~\cite{arvanitaki2016sound,Davis2024}. These systems have also enabled precision tests of Lorentz symmetry~\cite{lo2016acoustic,bushev2019testing,stanwix2005test} and have been proposed as probes of the quantum nature of gravity~\cite{guerreiro2022quantum,tobar2024detecting,cuthbertson1996parametric,hsiang2024graviton,aspelmeyer2022zeh}. In particular, proposed schemes for single-graviton detection using resonant-mass detectors require a ground-state-cooled mechanical resonator with a quadratic displacement readout~\cite{tobar2024detecting,tobar2025detecting}.

The importance of a purely quadratic readout is that it provides direct access to a displacement-squared observable \cite{brawley2016,Leijssen2017}. A conventional linear displacement readout measures the instantaneous position of the resonator. Since the position of a mechanical oscillator changes continuously during each oscillation cycle, a linear measurement is sensitive to the oscillator phase and can introduce measurement back-action that disturbs the mechanical energy. It therefore does not directly measure phonon number. In contrast, a purely quadratic readout measures a quantity proportional to $x^{2}$, which is insensitive to the sign of the displacement and is related to the energy stored in the mechanical mode. In the quantum regime, this energy is quantised in discrete phonon-number states. Therefore, if the linear term is sufficiently suppressed and the quadratic coupling is strong enough, the cavity-frequency shift can, in principle, provide information about phonon number without directly measuring the oscillator phase. This is the physical basis for a quantum non-demolition phonon-number measurement and motivates the development of the split-post geometry as a microwave platform for nonlinear opto-mechanical sensing.

In this work, we present a split-post re-entrant microwave resonator engineered to realise a \emph{quadratic} displacement readout. Re-entrant microwave cavities have previously served as highly sensitive \emph{linear} transducers~\cite{tobar1995characterizing,mcallister2017higher,parashar2024upconversion}, where the first-order frequency shift with displacement dominates the response. Here, the geometry and operating point are chosen to cancel the first-order sensitivity, $(df/dx=0)$, at a symmetry point, while retaining a finite second-order response, $(d^{2}f/dx^{2}\neq0)$. This yields a \emph{purely quadratic} response to leading order. Re-entrant cavities additionally offer strong parametric coupling~\cite{tobar1995characterizing,tobar2000accurate,locke1998parametric}, broadband tunability, and compatibility with ultra-low-noise detection, resulting in a versatile platform for quantum-limited sensing.

\section{\textbf{Theoretical Framework}}

The system under investigation comprises a mechanical membrane resonator enclosed within a microwave cavity incorporating a split-post geometry. This symmetry is analogous to that of a membrane-in-the-middle opto-mechanical system, where the membrane interacts with the electromagnetic field confined between two re-entrant posts (see Section \ref{QMCTD}). In general, to calculate the properties of such a microwave displacement transducer, we may express the cavity frequency $\omega_c$ as a Taylor expansion around the symmetry point $x=0$, as:
\begin{equation}
\omega_c(x) = \omega_c(0) + G_1 x + \frac{1}{2}G_2 x^2 +...
\label{parametricequation}
\end{equation}
Here $\omega_0$ is the unperturbed cavity resonant frequency with the membrane located at $x=0$. In general, allowing for a small displacement of the mechanical resonator, $\delta x$, The $n^{th}$-order optomechanical coupling coefficient is given by $G_n=\partial^n \omega_c/\partial x^n\rvert_{x=0}$. When the membrane's starting position is the cavity centre (symmetry line) we know that $\omega_c(x)=\omega_c(-x)$ and can therefore state that:
\begin{equation}
    G_1=\left.\frac{\partial \omega_c}{\partial x}\right\rvert_{x=0}=0,~~~G_2\neq0,
\end{equation}
implying an even function and quadratic coupling to the lowest order. In the off-centre position, we insert $x=x_0+\delta x$:
\begin{subequations}
\begin{align}
&\omega_c(x_0+\delta x)\approx\omega_c(0)+\frac{1}{2}G_2(x_0+\delta x)^2, \\
&=\omega_c(0)+\frac{1}{2}G_2x_0^2+G_2x_0\delta x+\frac{1}{2}G_2\delta x^2, \\
&\delta\omega_c=\omega_c(x_0)-\omega_c(x_0+\delta x) \\
&~~~~=-G_2x_0\delta x-\frac{1}{2}G_2\delta x^2.
\end{align}
\label{eq:Taylor}
\end{subequations}
We note that, in addition to the quadratic response, the system can also exhibit a linear dependence on displacement when the membrane is positioned asymmetrically, away from the central symmetry point. For a sinusoidal mechanical drive, the linear term produces a signal at the drive frequency, whereas the quadratic term produces a component at the second harmonic.

The dependence of the microwave frequency on the membrane{'}s displacement results in a standard opto-mechanical Hamiltonian description \cite{aspelmeyer2014cavity, burgwal2020comparing, jayich2008dispersive, thompson2008strong, brandao2020entanglement}:
\begin{subequations}
\begin{align}
\label{eq1}
&\hat{H} =\hat{H}_{MW}+\hat{H}_{mech}+\hat{H}_{int},\\
&\hat{H}_{MW}= \hbar \omega_{\text{c}} (\hat{a^{\dagger}}\hat{a}),\\
    % +\hat{a^{\dagger}}_{-}\hat{a}_{-}\right)\\
&\hat{H}_{mech} = \hbar \omega_{\text{m}} \left(\hat{b}^{\dagger}\hat{b}\right),\\
&\hat{H}_{int} = \hbar \left(g_{1} (\hat{b}+\hat{b}^\dagger) + g_{2}(\hat{b}+\hat{b}^\dagger)^2\right) (\hat{a}^{\dagger}\hat{a}).
    % -\hat{a^{\dagger}}_{-}\hat{a}_{-}\right) 
\end{align}
\end{subequations}

Here, we see that the system Hamiltonian is the sum of three sub-system Hamiltonians: a photon term representing the microwave cavity mode, $\hat{H}_{MW}$, a phonon term corresponding to the mechanical membrane mode, $\hat{H}_{mech}$, and an interaction term describing their mutual coupling, $\hat{H}_{int}$. In the above equations, $\hat{a}$ ($\hat{a}^\dagger$) represent the annihilation (creation) operators for microwave photons in the resonant cavity, whilst $\hat{b}$ ($\hat{b}^\dagger$) represent the annihilation (creation) operators for mechanical phonons in the membrane drum mode, which are related to the displacement operator by $\hat{x}=x_{zpf}(\hat{b}+\hat{b}^\dagger)$, where $x_{zpf}$ is the zero-point fluctuation displacement of the mechanical system (displacement created by single phonon occupancy). The coupling terms $g_1=G_2x_0x_{zpf}$ and $g_2=\frac{1}{2}G_2x_{zpf}^2$ represent the linear and quadratic opto-mechanical coupling rates. Thus, we can see in the $x_0=0$ symmetric case, the linear term $g_1$ vanishes and the interaction Hamiltonian is purely quadratic. 

If the cavity is strongly and coherently driven with the membrane in the symmetric case ($g_1=0$), we can state that $a=\alpha+\delta a$, $|\alpha|^2=\bar{n}_{cav}$, where $\alpha$ is the complex intracavity field amplitude and $\bar{n}_{cav}$ is the mean intracavity photon number. In this approximation, we can therefore write the interaction part of the Hamiltonian as:
\begin{equation}
    \hat{H}_{int}=\hbar g_2(|\alpha|^2+\alpha^*\delta a+\alpha\delta a^\dagger)(\hat{b}+\hat{b}^\dagger)^2,
\end{equation}
Here $\alpha^*$ is the complex conjugate of the field amplitude. Expanding the $(\hat{b}+\hat{b}^\dagger)^2$ term gives:
\begin{equation}
    \hat{H}_{int}=\hbar g_2(|\alpha|^2+\alpha^*\delta a+\alpha\delta a^\dagger)(1+2\hat{b}^\dagger \hat{b}+\hat{b}^2+\hat{b}^{\dagger 2}),
    \label{eq:expansion}
\end{equation}
which, when combined with $\hat{H}_{mech}$ and taking the rotating wave approximation (RWA), which allows the $\hat{b}^{\dagger 2}$ and $\hat{b}^2$ terms to be dropped, allows us to redefine the mechanical Hamiltonian as:
\begin{equation}
\hat{H}_{mech}^\prime\approx\hbar\left(\omega_m+2g_2|\alpha|^2\right)\hat{b}^\dagger\hat{b}.
\end{equation}
This is equivalent to the optical spring effect but originating from a quadratic coupling: the cavity photon population stiffens (or softens) the mechanical oscillator, hence changing its resonant frequency. 

Neglecting the constant frequency shift and the counter-rotating terms in eq. (\ref{eq:expansion}) we can now rewrite the system Hamiltonian as $\hat{H}=\hat{H}_{mech}^\prime+\hat{H}_{MW}+\hat{H}_{int}^\prime$, where
\begin{equation}
\hat{H}_{int}^\prime=2\hbar g_2(\alpha^*\delta a+\alpha\delta a ^\dagger)\hat{b}^\dagger\hat{b},
\end{equation}
which implies that microwave cavity fluctuations $\delta a$ are linearly coupled to the phonon number operator $\hat{b}^\dagger\hat{b}$. This permits a quantum non-demolition measurement of the phonon-number state (and hence mechanical resonator energy) via the cavity field phase, and is the primary reason a quadratic readout is desirable.

\section{Quadratic Microwave Cavity Transducer Design}
\label{QMCTD}

\begin{figure*}[t]
 \centering
        \includegraphics[width=0.67\textwidth]{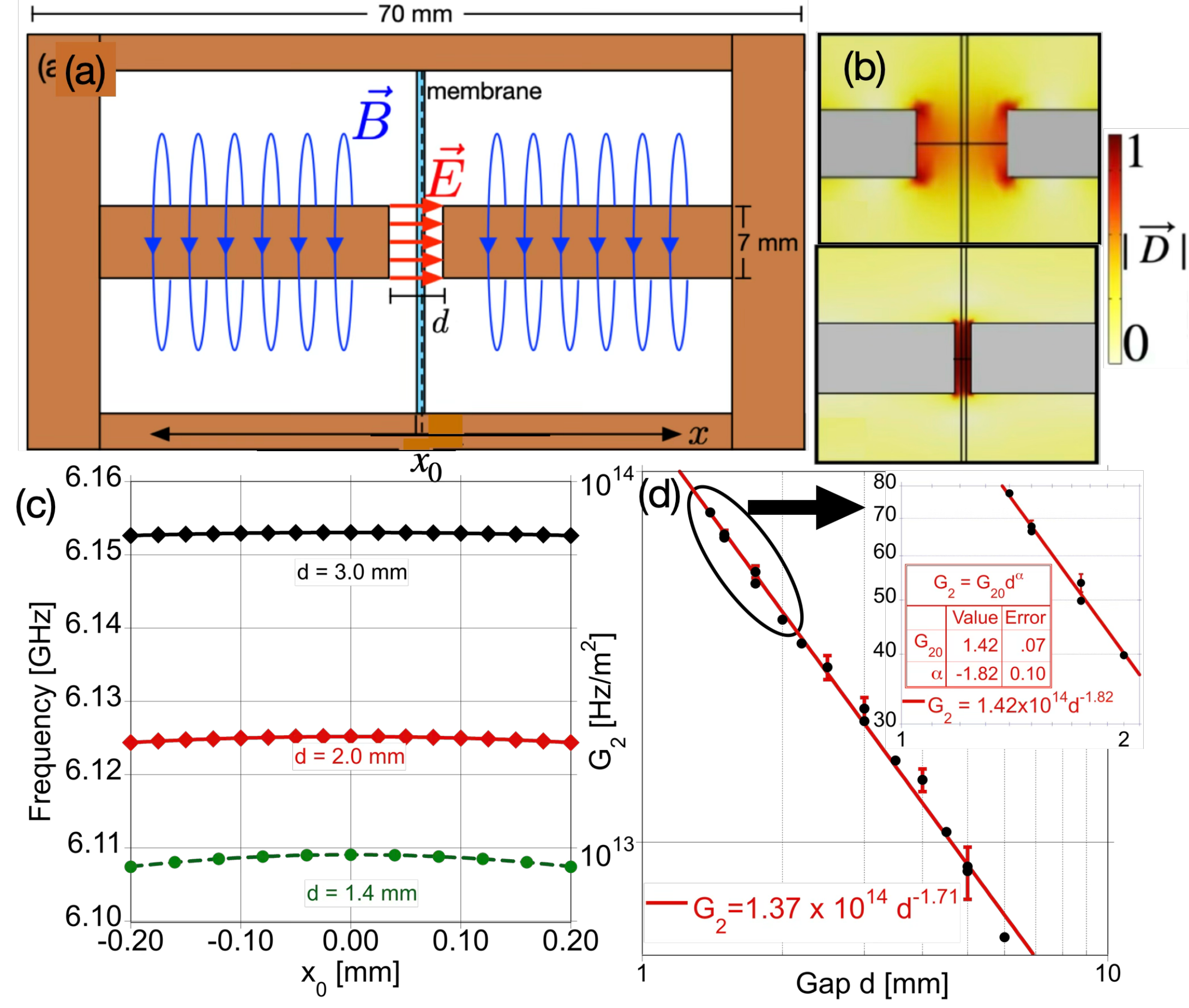}
\caption{Simplified schematic of the split-post re-entrant resonator with the sapphire acoustic membrane positioned at $x=x_0$. When $x_0=0$, the membrane is located at the central symmetry point and the linear coupling vanishes, giving a quadratic response as described by Eq.~(\ref{eq:Taylor}). The cavity electric field is localised between the post end faces, separated by the gap $d$, while the magnetic field $\vec{B}$ circulates around each post. The mechanical modes of the membrane modulate the microwave field through the change in membrane position. (b) Electric-field density from COMSOL simulations for gap spacings of $d=6.0$ (top) and $d=1.5 $ mm (bottom). This shows that, for large gap spacings, the mode transitions from a lumped 3D-LCR resonator to a cavity mode. (c) Simulated cavity resonance frequency as a function of membrane position $x_0$ for several modelled post gaps, $d$. The parabolic dependence about $x_0=0$ was used to extract the quadratic coupling coefficient $G_2$ for each gap spacing. (d) Extracted values of $G_2$ as a function of the gap $d$. For some gap spacings, more than one value was extracted using different mesh settings. We note small fluctuations between meshes, within the uncertainty of the fitted values of $G_2$. For the simulated values, the quadratic coupling increases strongly as the post gap was reduced, with the solid line showing a power-law fit of $G_2\propto d^{-1.7}$ over the range $d=1.4$--$6$ mm. Inset: over the reduced range $d=1.4$--$2$ mm, where the mode acts as a lumped 3D-LCR circuit, the extracted dependence is $G_2\propto d^{-1.83}$, with fitted exponent $-1.83\pm0.10$.}
\label{fig:topology}
\end{figure*}

The microwave cavity was specifically engineered to interrogate the acoustic modes of a sapphire membrane with a diameter of 50 mm and a thickness of 0.5 mm. The split-post geometry is advantageous because the re-entrant electric field is concentrated in the narrow region between the opposing post faces, where it strongly overlaps the sapphire membrane (see Fig. \ref{fig:topology}). The dielectric perturbation produced by membrane motion therefore modifies a significant fraction of the stored electric energy of the microwave mode, producing an enhanced displacement-to-frequency response. The cavity geometry was optimised using \textsc{COMSOL Multiphysics} simulations to support a re-entrant mode in the frequency range of $\sim 5-6$ GHz. The overall cavity length was set to 70~mm to achieve the desired microwave frequency while maintaining strong mode confinement and field localisation between the posts. 

To determine the value of $G_2$ (in units $Hz/m^2$) achievable in this system, we considered the static limit of Eq.~\ref{eq:Taylor}(a), with $\delta x=0$. In this case, the membrane position $x_0$ was varied statically, and the resulting cavity-frequency shift, $\delta f_c$ is quadratic in displacement. About the symmetric centre position, this shift is given by $\delta f_c=f_c(x_0)-f_c(0)=\frac{1}{2}G_2 x_0^2$. Thus, fitting the measured or simulated static frequency shift as a function of membrane position allows $G_2$ to be determined. Fig.~\ref{fig:topology}(a) shows the simplified resonator geometry, which acts as a 3D lumped-element resonator when the gap $d$ is sufficiently small \cite{mcallister2017higher}, with the electromagnetic field patterns indicated. We implemented this geometry in COMSOL to generate a finite-element mesh and simulate the frequency shift of the microwave resonator as the membrane position was varied. Fig.~\ref{fig:topology}(b) shows the modelled electric-field density, revealing that the device behaves as a three-dimensional lumped-element resonator for $d<2~\mathrm{mm}$, while for larger gaps it begins to transition towards a distributed cavity mode. Examples of the simulated frequency shift for several gap spacings, $d$, are shown in Fig.~\ref{fig:topology}(c). Fig.~\ref{fig:topology}(d) shows the fitted values of $G_2$ as a function of $d$ over the range $d=1.4$--$6.0~\mathrm{mm}$, with the inset showing a fit over the reduced range $d=1.4$--$2.0~\mathrm{mm}$.

\begin{figure*}[t]
\centering
\includegraphics[width=0.8\linewidth]{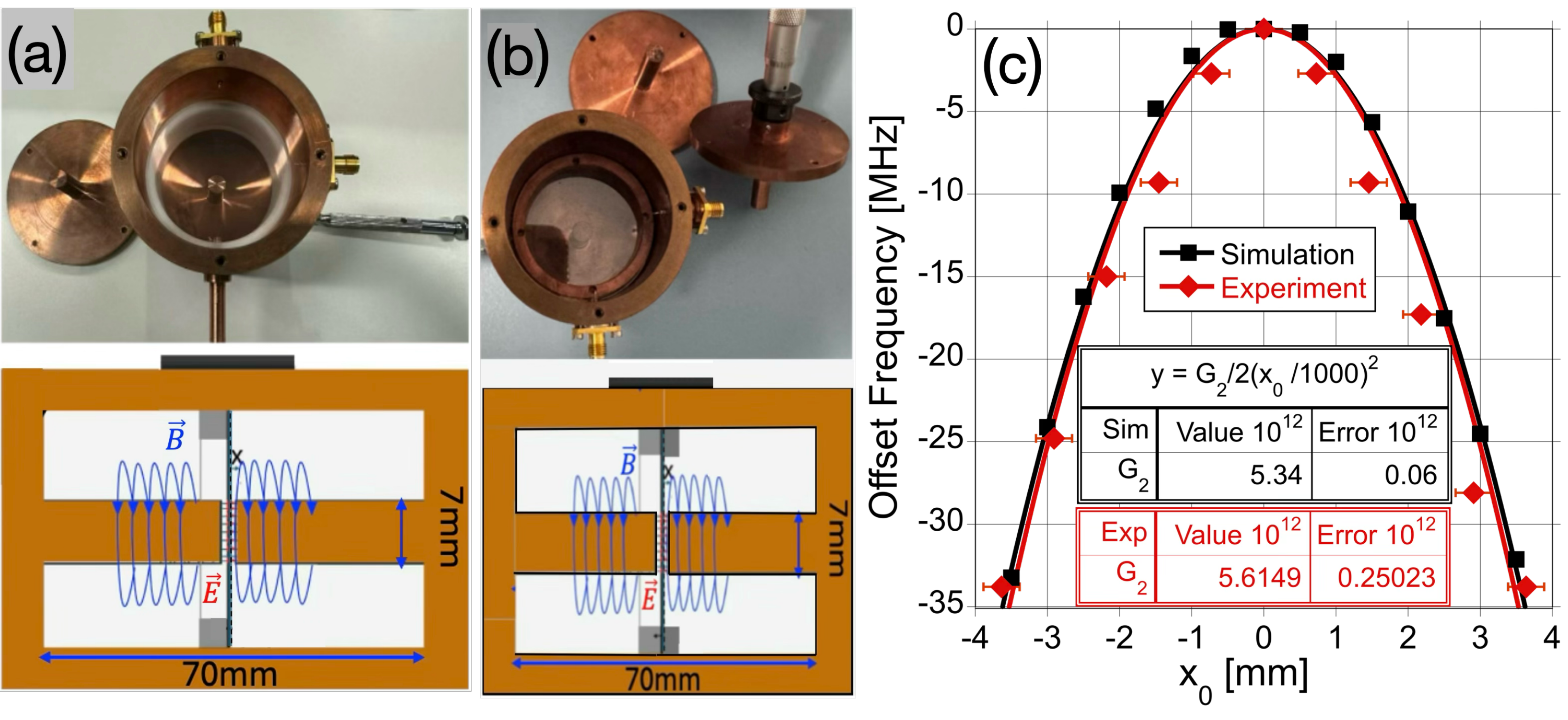}
\caption{Modified split-post microwave cavity resonators used to support the sapphire membrane. (a) To experimentally determine the static values of $G_2$, a Teflon ring was added to allow the membrane position, $x_0$, to be varied along the post gap while the microwave resonance frequency was measured using a vector network analyser (VNA). The upper image shows the interior of the fabricated cavity, while the lower image shows the corresponding simplified schematic, including the Teflon ring incorporated into the finite-element model. (b) For dynamic transducer operation, the cavity was modified to include a copper ring that held the membrane firmly above one of the posts. A tunable post mounted on a micrometer was then moved into position to maximise the $x^2$ response. (c) Measured and simulated values of $G_2$ for the cavity shown in (a) with $d=8~\mathrm{mm}$. The measured and simulated values are $(5.615\pm0.250)\times10^{12}~\mathrm{Hz/m^2}$ and $(5.341\pm0.055)\times10^{12}~\mathrm{Hz/m^2}$, respectively.}
\label{staticsetup} 
\end{figure*}

To demonstrate that the predicted values of $G_2$ could be realised experimentally, the split-post resonator was modified from the idealised geometry shown in Fig.~\ref{fig:topology}. A Teflon sliding support was fabricated to allow the membrane position, $x_0$, to be varied along the gap, as shown in Fig.~\ref{staticsetup}(a). A relatively large gap of approximately $8~\mathrm{mm}$ was required for practical manual tuning of the membrane position, since smaller gaps made it difficult to achieve the positioning precision needed to experimentally determine $G_2$. The microwave cavity response was characterised using a vector network analyser by measuring $S_{21}$ as $x_0$ was varied, allowing the resonance frequency and quality factor to be determined at each membrane position. The experimental results are plotted in Fig.~\ref{staticsetup}(c) and compared to COMSOL simulations.

To operate the cavity as an $x^2$ transducer, the sapphire membrane must be held firmly at the symmetric position, $x_0=0$. This was achieved using the cavity configuration shown in Fig.~\ref{staticsetup}(b), where a copper ring holds the membrane tightly against a Teflon ring positioned above one of the cavity posts. The top post was adjusted with a micrometer to produce a small gap of $d=1.5~\mathrm{mm}$, allowing the acoustic modes of the membrane to modulate the microwave mode and operate as a quadratic displacement sensor. To operate the cavity as a linear displacement sensor, the top post was tuned to increase the gap spacing, placing the membrane away from the symmetric position. In this configuration, the response is better described by the linear term in Eq.~(\ref{eq:Taylor}c). These measurements are described in the next section.

\begin{figure}[h]
\centering
\includegraphics[width=1.0\columnwidth]{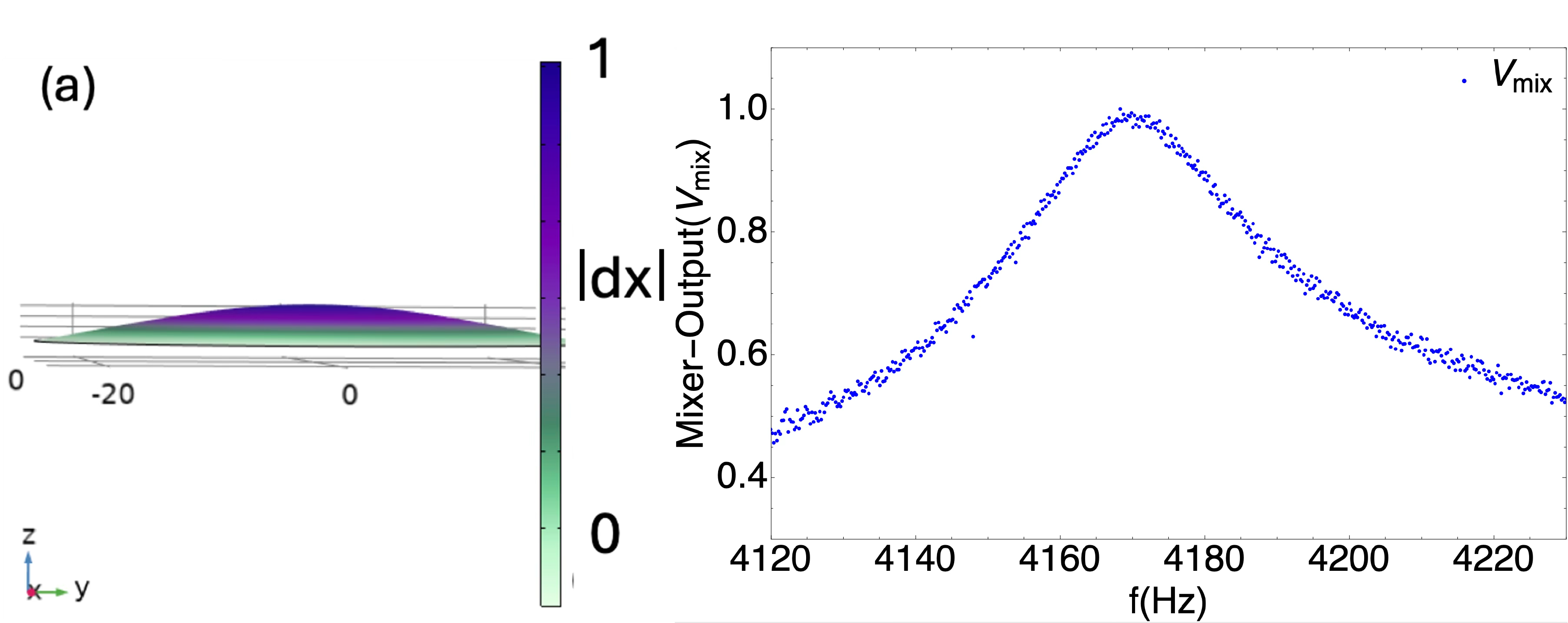}
\caption{\label{fig:membranedisplacement} 
{The mechanical resonance of the sapphire membrane: (a) simulation of the fundamental drum mode, showing the crystal{'}s absolute displacement. (b) The measured frequency response of this resonance at {$f_m=4.169$ kHz}. The mode is excited using a PZT drive and read via the mixer output of the interferometer.}}
\end{figure}

The mechanical properties of the sapphire membrane were also modelled using \textsc{COMSOL} to determine the membrane-mode resonant frequencies, $f_m$. The outer boundary of the membrane was clamped, and the eigenfrequencies, modal effective masses, and participation factors of the drum modes were evaluated. The fundamental drum-mode resonant frequency was calculated to be $f_m=\omega_m/2\pi=4.169~\mathrm{kHz}$, with a modal effective mass of $m_m=2.7~\mathrm{g}$. Sweeping the PZT actuator drive through this frequency and monitoring the interferometer mixer output experimentally confirmed the presence of this mechanical mode and its coupling to the microwave resonator. Fig.~\ref{fig:membranedisplacement} shows (a) the simulated absolute displacement, denoted by $|\mathrm{d}x|$, of the fundamental drum mode and (b) the mixer output voltage, $V_{\mathrm{mix}}$, measured while driving the fundamental drum mode.

The measured resonance shown in Fig. \ref{fig:membranedisplacement} (b) has an approximate full width at half maximum of $\Delta f_{\mathrm{FWHM}}\simeq50~\mathrm{Hz}$. The corresponding mechanical quality factor is therefore $Q_m=f_m/\Delta f_{\mathrm{FWHM}}\simeq83$. Equivalently, the mechanical energy-decay linewidth is $\Gamma_m/2\pi\simeq50~\mathrm{Hz}$. The relatively modest value of $Q_m$ is attributed to the room-temperature clamping and mechanical support required to couple the PZT drive to the membrane.

\section{Dynamic Microwave Readout of Linear and Quadratic Membrane Motion}

\begin{figure}[h]
\centering
\includegraphics[width=1.0\linewidth]{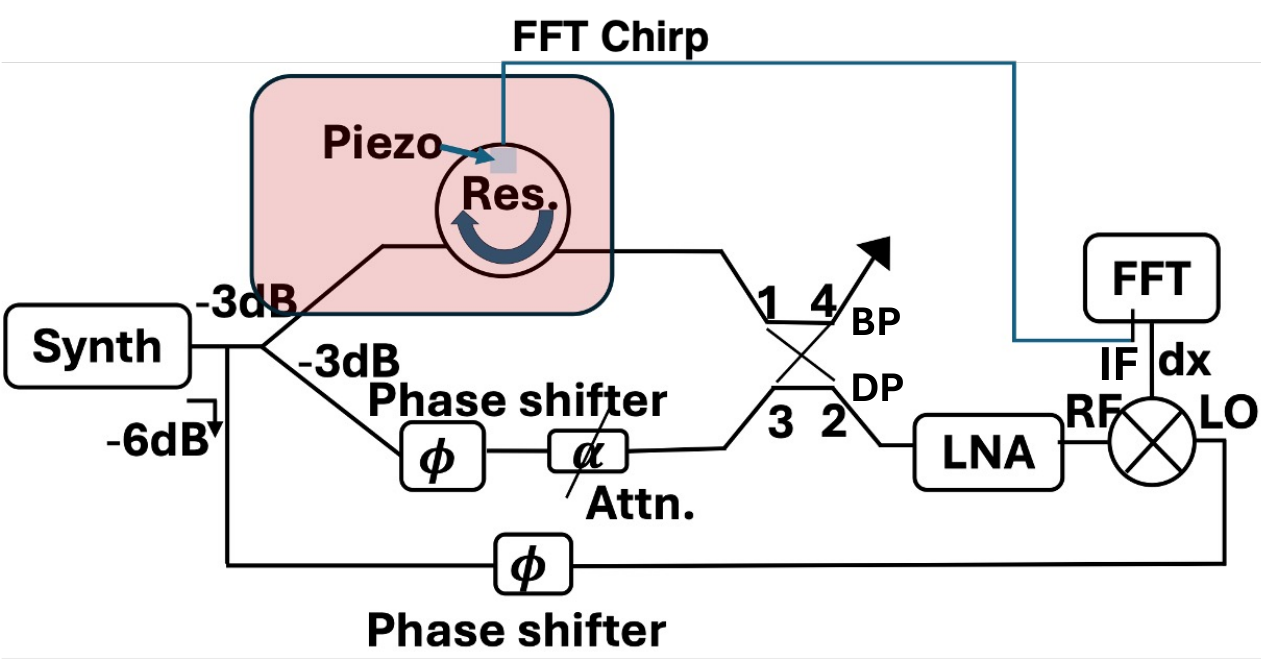}
\caption{\label{fig:experimentalsetup} 
{Experimental setup used to detect motion of the mechanical membrane mode within the split-post microwave resonator. The interferometer is balanced to minimise the carrier at the dark port (DP), strongly suppressing unwanted carrier-related signals. A spectrum analyser operating in FFT mode was used both to drive the piezoelectric actuator and to measure the interferometer output. The interferometer frequency-to-voltage sensitivity, $\partial V_{\rm mix}/\partial f_c$, is maximised when the carrier is suppressed at the DP. At this operating point, flicker noise from the low-noise amplifier (LNA) is also suppressed, and the dominant noise contribution arises from synthesiser phase noise converted by the dispersive response of the resonator. Introducing a matched dispersive element in the reference arm would make this source noise predominantly common mode, after which the measurement floor would be set primarily by the noise temperature of the first-stage amplifier and the microwave loss preceding it.}}
\end{figure}

The experimental setup implemented to observe the motion of the mechanical membrane mode inside the split-post microwave resonator is shown in Fig.~\ref{fig:experimentalsetup}. It is based on a low-noise precision interferometric readout \cite{ivanov1998microwave}, which converts membrane-induced cavity-frequency shifts into a voltage at the mixer output. One arm of the interferometer contains the optomechanical cavity transducer, whose microwave resonance frequency is modulated by the motion of the membrane. The other arm acts as a reference, with adjustable phase and attenuation. When the two arms are recombined, the output voltage is sensitive to the phase difference between the reference and cavity arms, providing a homodyne readout of the membrane-induced cavity-frequency modulation.

We characterise the interferometer sensitivity to shifts in the microwave resonance frequency by the quantity ${\partial V_{\rm mix}}/{\partial f_c}$. This sensitivity depends on the microwave coupling to the resonant mode, the loaded microwave $Q$-factor, the incident power, the interferometer balance, and the mixer conversion efficiency. It is calibrated by applying a known frequency modulation to the microwave source, which mimics a membrane-induced cavity-frequency shift, and measuring the corresponding mixer output $\partial V_{\rm mix}$ at the modulation frequency for a known modulation amplitude $\partial f_c$.

A key feature of the interferometric readout is the use of the dark port (DP) of the hybrid coupler. The microwave carrier is divided between the test arm, containing the resonator, and the balance arm, containing a phase shifter and attenuator. These components are adjusted so that the two large carrier fields recombine destructively at the DP. The dark port therefore suppresses the unmodulated carrier at the cavity resonance frequency, while the small phase-modulation sidebands produced by membrane motion remain available for detection. This is important because the carrier power is much larger than the displacement-induced sidebands. Without carrier suppression, the large carrier would drive the low-noise amplifier (LNA) into compression or saturation, forcing operation at reduced microwave power and degrading the readout sensitivity. By nulling the carrier at the DP, the LNA can be placed after the interferometer and used to amplify the small motion-induced sidebands without being saturated by the carrier \cite{ivanov1998microwave,regal2008measuring, woode1998application}, so the sensitivity is increased by the gain of the LNA. The amplified signal is then mixed down with the local oscillator (LO), whose phase is adjusted so that the mixer IF voltage is proportional to the phase quadrature of the RF signal. The mixer output is finally sent to a spectrum analyser or fast Fourier transform (FFT) machine, allowing phase changes due to membrane-induced variations of the microwave cavity frequency to be measured.

The interferometric readout operated with a maximum sensitivity of about $\partial V_{\rm mix}/\partial f_c\sim 4.5\times10^{-6}$ $\mathrm{V/Hz}$, although this sensitivity varied with the membrane position, $x_0$, within the microwave cavity. Changing the gap spacing of the split-post resonator altered both the resonant frequency and loaded quality factor of the microwave mode, owing to the corresponding variation in loss and mode coupling, and therefore changes the sensitivity of the interferometric readout. Excitation of the membrane mode by the piezoelectric transducer produced a displacement-induced phase shift, or equivalently a frequency shift, in the microwave signal, which is measured at the IF port of the mixer. The microwave cavity and PZT actuator were housed inside a vacuum chamber maintained at a pressure of $1\times10^{-5}~\mathrm{mbar}$ at room temperature. The FFT analyser recorded the mixer output while simultaneously driving the PZT with its internal chirp source followed by a low-pass filter.

The interaction between the membrane displacement, $\delta x$, and the piezoelectric drive voltage, $\delta V_{\rm PZT}$, is expected to be approximately linear only in the small-signal regime used in this experiment. In this limit, the applied voltage produces a proportional strain in the PZT actuator, which is mechanically transferred to the sapphire membrane through a mode-dependent transfer function. At angular frequency $\omega$, this response can be written as
\begin{equation}
\label{equation10}
    \delta x(\omega)=S(\omega)R(\omega)\delta V_{\rm PZT}
    =T(\omega)\delta V_{\rm PZT},
\end{equation}
where $R(\omega)$ is the voltage-to-displacement response of the PZT actuator, $S(\omega)$ is the mechanical transfer function between the actuator and the membrane mode, and $T(\omega)$ is the combined transfer function. For a fixed excitation frequency and sufficiently small drive amplitude, $T(\omega)$ is approximately constant, giving a linear relationship between $\delta x$ and $\delta V_{\rm PZT}$.

We use a PA4FKH3 piezoelectric actuator with a no-load displacement of order $\sim20~\mathrm{nm/V}$ for $\delta V_{\rm PZT}\leq 1~\mathrm{V}$. This linear response does not persist at large drive amplitudes, up to the rated maximum of $150~\mathrm{V}$, where PZT hysteresis and nonlinear actuator behaviour become significant. Furthermore, we expect that the specified no-load displacement is not transferred directly to the membrane because of imperfect mechanical clamping and mass loading by the cavity structure. On the other hand the acoustic Q-factor of the mode will enhance the resonant response by a factor related to $Q_m$, which for our experiment is of order 50-100. We observe no evidence of nearby parasitic mechanical resonances in the support structure, which could otherwise cause both the magnitude and phase of $T(\omega)$ to vary rapidly with frequency. For the measurements presented here, the drive amplitude was maintained within the small-signal regime, and the membrane resonance showed no interaction with nearby parasitic modes.

The mixer output response then depends on the operating position of the membrane within the cavity. At the symmetric point, $x_0=0$, the linear coupling is suppressed and the cavity response is dominated by the quadratic term in Eq.~(\ref{eq:Taylor}). Thus, for a sinusoidal membrane displacement at the acoustic frequency $f_m$, the microwave signal appears at the second harmonic, $2f_m$, with an amplitude proportional to $\delta x^2$, and hence to $\delta V_{\rm PZT}^2$. In contrast, when the membrane is offset from the symmetry point, $x_0\neq0$, a linear coupling term is introduced. In this configuration, the mixer output occurs primarily at the acoustic frequency $f_m$, with an amplitude proportional to $\delta x$, and hence to $\delta V_{\rm PZT}$.

Given that the shift in microwave cavity frequency for the first and second harmonic is given by $\delta f_{c1}=-G_2x_0\delta x$ and $\delta f_{c2}=-\frac{G_2}{2}\delta x^2$ respectively, and defining the conversion ratio of the frequency discriminator setup as $K_{FD}=\frac{\partial V_{mix}}{\partial{f_c}}$, then combing this with Eq. (\ref{equation10}), gives the following change in mixer output voltage at the first and second harmonic respectively;
\begin{equation}
\begin{aligned}
\delta V_{\mathrm{mix},1}
    &= K_{\mathrm{FD},1}G_{2,1}x_0T_1\delta V_{\mathrm{PZT}}, \\
\delta V_{\mathrm{mix},2}
    &= K_{\mathrm{FD},2}G_{2,2}T_2^2\delta V_{\mathrm{PZT}}^2.
\end{aligned}
\label{eq3}
\end{equation}
Here, $G_{2,2}$ is the value of $G_2$ when $x_0=0$ for gap spacing, $d_2$, and $G_{2,1}$, is the value of $G_2$ when $x_0 \ne 0$ for gap spacing $d_1$.

\section{Results and Discussion} 

\subsection{Demonstrator Experiment}

To experimentally characterise the dynamic performance of the split-post transducer, the cavity was configured as shown in Fig.~\ref{staticsetup}(b) and integrated with the interferometric readout shown in Fig.~\ref{fig:experimentalsetup}. The dynamic response was measured by driving the sapphire membrane with a PZT actuator and detecting the resulting modulation of the microwave cavity resonance. A chirp signal was applied to the PZT, with the drive voltage varied from $10~\mathrm{mV}$ to $1000~\mathrm{mV}$, over a frequency range centred on the acoustic membrane mode at $f_m=4.169~\mathrm{kHz}$. The peak mixer output, $V_{\rm mix}$, was measured using a spectrum analyser and recorded as an RMS voltage.

\begin{figure*}[t]
\centering
\includegraphics[width=0.8\linewidth]{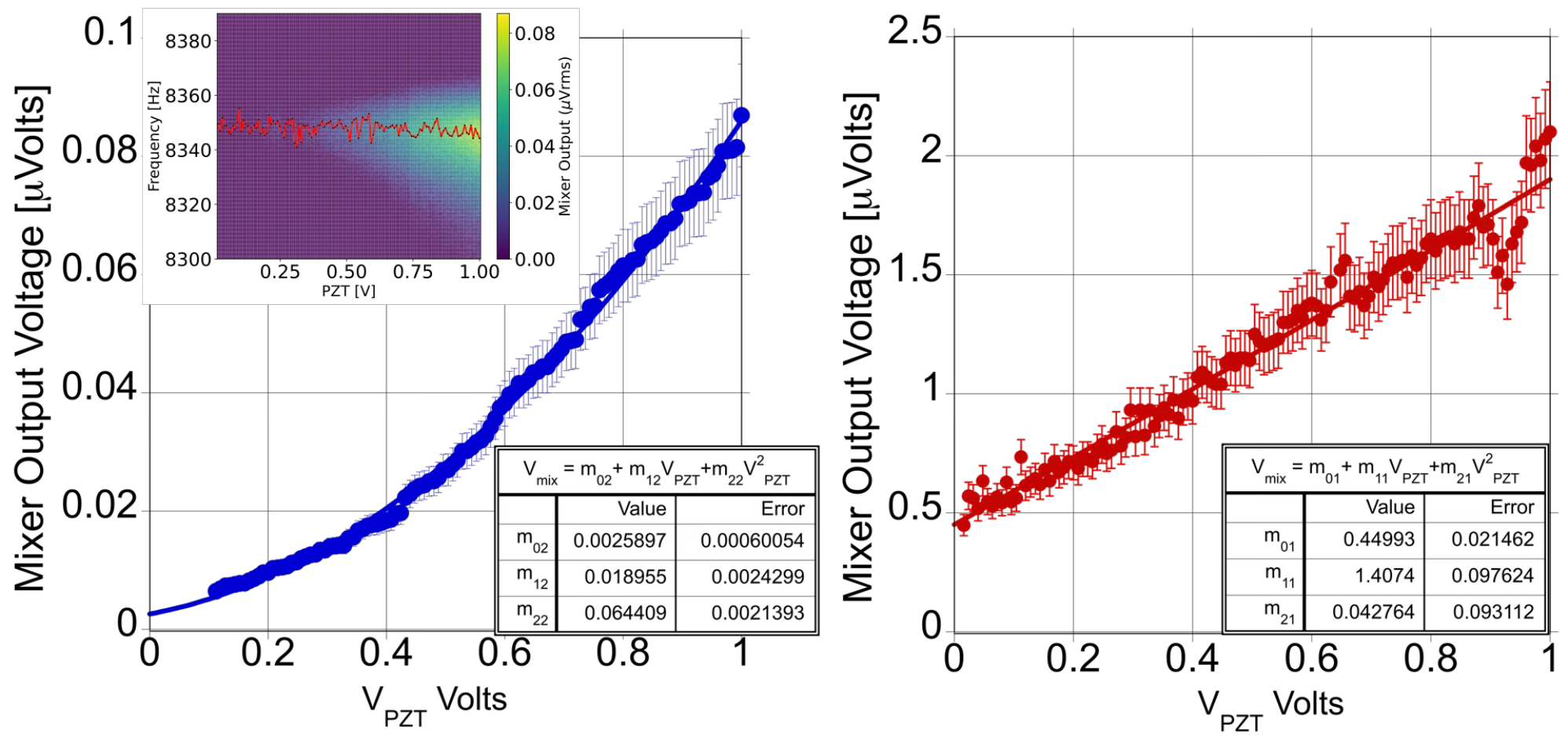}
\caption{Mixer output voltage, $V_{\rm mix}$, as a function of the applied piezoelectric drive voltage, $V_{\rm PZT}$, for dynamic excitation of the sapphire membrane mode. The PZT was driven with a chirp signal centred on the fundamental acoustic drum-mode frequency, $f_m$, shown in Fig.~\ref{fig:membranedisplacement}. Left: response with the membrane positioned at the central symmetry point, $x_0=0$ with $d_2=1.5~\mathrm{mm}$, where the linear coupling is suppressed and the dominant signal occurs at the second harmonic, $2f_m$. A second-order polynomial fit shows that the measured peak RMS mixer voltage has a dominant quadratic dependence on $V_{\rm PZT}$, with a fitted $V_{\rm PZT}^2$ coefficient of $m_{22}=0.0644\pm0.0021~\mu\mathrm{V}/\mathrm{V}^2$. The inset shows a density plot of the mixer output around $2f_m$, with drive voltage on the horizontal axis, frequency on the vertical axis, and colour indicating the mixer output voltage. The red trace follows the locus of the maximum response as the drive voltage was varied. Right: first-harmonic response measured with the membrane displaced by $3.5~\mathrm{mm}$ from the symmetry point after increasing the gap to $d_1=8~\mathrm{mm}$. In this configuration, the dominant signal occurs at the fundamental acoustic frequency, $f_m$. The second-order polynomial fit shows that the linear term dominates, with a fitted coefficient of $m_{11}=1.407\pm0.093~\mu\mathrm{V}/\mathrm{V}$.}
\label{fig:Offcenterandcenterposition} 
\end{figure*}

The membrane was first held at the centre of a gap of $d_2=1.5~\mathrm{mm}$. Since the membrane thickness is $0.5~\mathrm{mm}$, this left a clearance of $0.5~\mathrm{mm}$ on either side of the membrane. In this symmetric configuration, the linear coupling was suppressed and the quadratic response was measured at the second harmonic, $2f_m\sim8.34~\mathrm{kHz}$. The top post was then retracted to increase the gap to $d_1=8~\mathrm{mm}$, placing the membrane approximately $3.5~\mathrm{mm}$ away from the centre of the gap. In this offset configuration, the response was dominated by linear coupling, and the first-harmonic signal at $f_m$ was measured. The results are plotted against the applied excitation voltage to the piezoelectric crystal ($V_{PZT}$)in Fig.~\ref{fig:Offcenterandcenterposition}.

The voltage-to-displacement transfer function, $T(f)$, may be estimated from the fitted first- and second-harmonic responses in Fig.~\ref{fig:Offcenterandcenterposition}. From Eq.~\ref{eq3}, the fitted coefficients satisfy $K_{\rm FD1}G_{2,1}x_0T_1(f_m)=m_{11}$ and $K_{\rm FD2}G_{2,2}T_2^2(f_m)=m_{22}~\mathrm{V}^{-1}$, where $K_{\rm FD1}$ and $K_{\rm FD2}$ are the frequency-to-voltage conversion coefficients for the two measurement configurations. For the off-centred configuration with $d_1=8~\mathrm{mm}$ and $x_0=3.5~\mathrm{mm}$, the quadratic coupling coefficient was determined from the static measurements to be $G_{2,1}=(5.61\pm0.25)\times10^{12}~\mathrm{Hz/m^2}$. For the centred configuration with $d_2=1.5~\mathrm{mm}$ and $x_0=0$, simulations gave $G_{2,2}\simeq7\times10^{13}~\mathrm{Hz/m^2}$. For the centred configuration, the dynamic displacement response was estimated from $T_2=\sqrt{m_{22}/(K_{\rm FD2}G_{2,2})}$, where $m_{22}=(0.0644\pm0.0021)\times10^{-6}~\mathrm{V}^{-1}$. Substitution of the measured and simulated parameters gives $T_2\sim14~\mathrm{nm/V}$. The post was then detuned to obtain the off-centred configuration with $d_1=8~\mathrm{mm}$, for which the predominantly linear response gives $T_1=m_{11}/(K_{\rm FD1}G_{2,1}x_0)$, where $m_{11}=(1.407\pm0.093)\times10^{-6}$. The frequency-to-voltage conversion was reduced to $0.1\times10^{-6}$ $\mathrm{V/Hz}$, in this configuration, giving $T_1\sim1~\mathrm{nm/V}$. These estimates provide a sanity check showing that the inferred membrane displacements are physically reasonable. A more precise determination of $T(f)$ would require an independent in situ calibration of both the mechanical transfer function and the microwave frequency-to-voltage conversion. However, the next stage of this work will employ high-$Q$ acoustic and microwave resonators operating at millikelvin temperatures. In this regime, it should be possible to parametrically excite and calibrate the transducer in situ, eliminating the need for the PZT actuator and avoiding the excess mechanical support losses present in the demonstrator system, which are required to make the membrane sensitive to the PZT.

From our experiments we may characterise the acoustic system, and assess what could be achieved in the future. Comparing the results shown in Eq. (\ref{parametricequation}) with the fitted value of $G_2$ in Fig.~\ref{fig:topology}, the extracted quadratic coupling coefficient, $G_2$ when placed in the centre of the cavity is $G_2=7\times10^{13}~\mathrm{Hz/m^2}$. The zero-point fluctuation amplitude of the membrane mode, $x_{\mathrm{zpf}}=\sqrt{\frac{\hbar}{2m_{\mathrm{eff}}\omega_m}}$, is calculate to be, $x_{\mathrm{zpf}}=8.6\times10^{-19}~\mathrm{m}$, and thus a single-photon second-order opto-mechanical coupling rate of, $g_2=\frac{1}{2}G_2x_{\mathrm{zpf}}^2 \simeq 2.6\times10^{-23}~\mathrm{Hz}$. At room temperature, $T=295~\mathrm{K}$, the thermal phonon occupation of the membrane mode is $n_{\mathrm{th}} =\frac{1}{\exp(\hbar\omega_m/k_B T)-1} \simeq$  $\frac{k_B T}{\hbar\omega_m} \simeq 1.5\times10^{9}$, for $f_m=4.169~\mathrm{kHz}$, confirming that the present measurement is in the classical thermal regime. The corresponding single photon quadratic cooperativity is, $C_2^{(0)}=\frac{4g_2^2}{\kappa\Gamma_m}$, here $\kappa$ is the microwave cavity linewidth, and $\Gamma_m$ is the mechanical linewidth. For the present room-temperature experiment, $\kappa\simeq3.5\times10^6~\mathrm{Hz}$, and the measured mechanical linewidth is approximately $\Gamma_m\simeq50~\mathrm{Hz}$, giving, $C_2^{(0)}\simeq1.5\times10^{-53}$. This value is far below unity, confirming that the present room-temperature experiment is a classical demonstration of tunable quadratic transduction rather than an already quantum-limited transducer.

\subsection{Future Improvements}

Reaching appreciable cooperativity requires a redesigned device rather than cryogenic operation alone. Under microwave driving, the quadratic cooperativity is $C_2=4g_2^2n_c/(\kappa\Gamma_m)=n_cC_2^{(0)}$, where $n_c$ is the mean intracavity photon number and all coupling rates and linewidths are expressed in Hz. Taking as a representative future target a superconducting cavity with $Q_c\sim10^7$ and a mechanical mode with $Q_m\sim10^8$, the corresponding linewidths are $\kappa=f_c/Q_c\simeq530~\mathrm{Hz}$ and $\Gamma_m=f_m/Q_m\simeq4.2\times10^{-5}~\mathrm{Hz}$ for $f_c\simeq5.3~\mathrm{GHz}$ and $f_m=4.169~\mathrm{kHz}$. Unity driven cooperativity therefore requires $g_2\sqrt{n_c}=\sqrt{\kappa\Gamma_m/4}\simeq7.4\times10^{-2}~\mathrm{Hz}$. Even for a very large intracavity photon number of $n_c\sim10^{12}$, this corresponds to a required single-photon quadratic coupling of $g_2\simeq7.4\times10^{-8}~\mathrm{Hz}$. Compared with the present estimate of $g_2\simeq2.6\times10^{-23}~\mathrm{Hz}$, this represents an increase of approximately $2.9\times10^{15}$. 

Closing this gap requires reducing the effective modal mass, increasing the electromagnetic curvature $G_2$, or both. Taking the re-entrant-cavity scaling $G_2\propto d^{-1.8}$, adopting a mechanical resonator comparable to that of Ref.~\cite{jayich2012}, with $f_m=261~\mathrm{kHz}$, $\Gamma_m=65~\mathrm{mHz}$, and $m_{\mathrm{eff}}\approx75~\mathrm{ng}$, and simultaneously reducing the post gap from $1.5~\mathrm{mm}$ to approximately $1~\mu\mathrm{m}$ would increase $g_2$ by a factor of approximately $3\times10^{11}$. Starting from the present estimate $g_2\simeq2.6\times10^{-23}~\mathrm{Hz}$, this gives $g_2\simeq8\times10^{-12}~\mathrm{Hz}$, comparable to the order-of-magnitude estimate of $10^{-11}~\mathrm{Hz}$. This represents a substantial improvement, although it remains approximately four orders of magnitude below the target value $g_2\simeq7.5\times10^{-8}~\mathrm{Hz}$. This estimate uses the scaling $G_2\propto d^{-1.8}$ extracted from the present cavity topology, in which the gap was reduced while all other geometric parameters, including the post width and profile, cavity radius, and cavity height, were held fixed at values selected for the current unoptimised design rather than for micron-scale operation. Since the post geometry and cavity aspect ratio also determine the field concentration and curvature obtained for a given gap, re-optimising these parameters specifically for micron-scale operation may increase $G_2$, and hence $g_2$, beyond the simple gap scaling considered here. Closing the remaining gap to $g_2\simeq7.5\times10^{-8}~\mathrm{Hz}$ should therefore be regarded as a combined geometry- and mass-optimisation problem, with full electromagnetic optimisation of the re-entrant topology at small gap being a priority for future device development.

A comparable redesign is required to bring the system into the resolved-sideband regime, a separate but related prerequisite for sideband cooling, strong dynamical back-action, and phonon-number-resolved measurements. The present microwave resonance sits at $f_c\simeq6.1~\mathrm{GHz}$ with a measured loaded quality factor $Q_l\simeq1.5\times10^3$, giving $\kappa=f_c/Q_l\simeq4.1~\mathrm{MHz}$, far larger than the membrane frequency $f_m=4.169~\mathrm{kHz}$ ($f_m\ll\kappa$). Resolved-sideband operation instead requires $\kappa<f_m$, equivalently $Q_l>f_c/f_m\simeq1.5\times10^6$ for the present mode, roughly a thousand-fold improvement on the room-temperature prototype. This is not a further demand beyond the cooperativity target above but the same underlying requirement, a low-loss superconducting cavity and reduced mechanical damping: loaded quality factors as high as $5\times10^8$ have already been demonstrated in re-entrant cavities \cite{bassan2008parametric}, comfortably exceeding what either condition needs. The requirement is further relaxed by moving to higher-frequency acoustic modes; a $1~\mathrm{MHz}$ mechanical mode at the same microwave frequency would need only $Q_l>6.1\times10^3$, easing both the sideband-resolution and thermal-occupation constraints simultaneously.

Taken together, these estimates point to a common set of design changes: a superconducting cavity, cryogenic operation, a lower-mass and higher-frequency mechanical mode, and, as argued above, a re-entrant geometry re-optimised for small gaps. Pursuing $g_2$ enhancement through geometry and mass, and pursuing sideband resolution through $Q_l$ and $f_m$, are therefore complementary parts of a single upgrade path towards quantum-regime operation of the split-post transducer, consistent with related sapphire-based microwave acoustic transducer platforms \cite{cuthbertson1996parametric}.

\subsection{Quadratic Sensitivity and Readout Noise}

At the symmetric operating point, the transducer is intentionally quadratic, and a conventional linear displacement sensitivity in $\mathrm{m}/\sqrt{\mathrm{Hz}}$ is therefore not the most natural figure of merit because $G_1=df_c/dx=0$. The cavity-frequency shift is instead given by $\Delta f_c=\frac{1}{2}G_2x^2$. The mixer-voltage noise amplitude spectral density, $\sqrt{S_V}$, may first be converted into an equivalent cavity-frequency noise according to $\sqrt{S_f}=\sqrt{S_V}/|\partial V_{\mathrm{mix}}/\partial f_c|$. The corresponding noise-equivalent displacement-squared sensitivity is then $\sqrt{S_{x^2}}=2\sqrt{S_f}/|G_2|$, with units of $\mathrm{m^2}/\sqrt{\mathrm{Hz}}$, and provides the appropriate sensitivity measure at the quadratic operating point \cite{brawley2016,Leijssen2017}.

In the present interferometric readout, the dominant technical noise is associated with the phase noise of the microwave synthesiser shown in Fig.~\ref{fig:experimentalsetup}, which has a one-sided phase-noise power spectral density of $S_{\phi,\mathrm{osc}}\simeq2.0\times10^{-11}~\mathrm{rad^2/Hz}$, or $\sqrt{S_{\phi,\mathrm{osc}}}\simeq4.5\times10^{-6}~\mathrm{rad}/\sqrt{\mathrm{Hz}}$. In the unresolved-sideband limit, the dispersive cavity arm and the nondispersive reference arm do not experience identical phase fluctuations. Using the cavity phase slope $2Q_L/f_c$, the residual differential phase-noise amplitude may be approximated by $\sqrt{S_{\phi,\mathrm{res}}(f_m)}=(2Q_Lf_m/f_c)\sqrt{S_{\phi,\mathrm{osc}}(f_m)}$. Referring this residual noise back to an equivalent cavity-frequency fluctuation gives $\sqrt{S_f(f_m)}=\sqrt{S_{\phi,\mathrm{res}}}/(2Q_L/f_c)=f_m\sqrt{S_{\phi,\mathrm{osc}}(f_m)}$. The corresponding oscillator-limited quadratic sensitivity is therefore $\sqrt{S_{x^2}(f_m)}=2f_m\sqrt{S_{\phi,\mathrm{osc}}(f_m)}/|G_2|$. Using $f_m=4.169~\mathrm{kHz}$ and the predicted quadratic curvature $G_2\simeq7\times10^{13}~\mathrm{Hz/m^2}$ for the centred $1.5~\mathrm{mm}$-gap configuration, and the synthesiser phase noise, gives a displacement-squared sensitivity of approximately $\sqrt{S_{x^2}}\simeq5.3\times10^{-16}~\mathrm{m^2}/\sqrt{\mathrm{Hz}}$. Replacing the present source with a state-of-the-art oscillator having $S_{\phi,\mathrm{osc}}(f_m)\simeq 10^{-16}~\mathrm{rad^2/Hz}$ would reduce this contribution to approximately $\sqrt{S_{x^2}}\simeq1.2\times10^{-18}~\mathrm{m^2}/\sqrt{\mathrm{Hz}}$.

Further rejection of synthesiser phase noise could be achieved by placing a matched dispersive element in the reference arm, causing the source fluctuations to become predominantly common mode. Once this contribution is suppressed, the sensitivity would be limited by the added noise and losses of the measurement chain, particularly the noise temperature of the first-stage LNA and the microwave loss preceding it. 

Once the microwave-source phase noise is suppressed using matched dispersive elements in the two interferometer arms, the readout sensitivity is limited by the noise temperature, $T_N$, of the first-stage microwave amplifier. For a carrier power $P_c$ incident on the amplifier, the corresponding phase-noise spectral density referred to the interferometer output is approximately $S_{\phi}^{\mathrm{amp}}=k_BT_N/P_c$. Using the cavity phase response, $\partial\phi/\partial f_c=2Q_L/f_c$, this gives an equivalent cavity-frequency noise of $\sqrt{S_f^{\mathrm{amp}}}=(f_c/2Q_L)\sqrt{k_BT_N/P_c}$. The resulting amplifier-limited displacement-squared sensitivity is therefore $\sqrt{S_{x^2}^{\mathrm{amp}}}=(f_c/|G_2|Q_L)\sqrt{k_BT_N/P_c}$. Thus, the sensitivity improves with increasing carrier power, cavity quality factor, and quadratic curvature, and scales as $\sqrt{S_{x^2}^{\mathrm{amp}}}\propto\sqrt{T_N}$. Replacing the room-temperature amplifier with a quantum-limited superconducting parametric amplifier would consequently provide a direct improvement in the quadratic sensitivity. For comparison, displacement-squared sensitivities of $3.3\times10^{-24}~\mathrm{m^2}/\sqrt{\mathrm{Hz}}$ and $2.0\times10^{-25}~\mathrm{m^2}/\sqrt{\mathrm{Hz}}$ have been demonstrated in nanoscale optical systems \cite{brawley2016,Leijssen2017}; however, differences in modal mass, mechanical frequency, quadratic coupling, and readout architecture prevent a direct comparison with the present macroscopic microwave transducer.

The quadratic sensitivity may also be expressed as an energy or phonon-number sensitivity. For a harmonic mechanical mode, the cycle-averaged energy is $E=m_{\mathrm{eff}}\omega_m^2\langle x^2\rangle$, giving $\sqrt{S_E}=m_{\mathrm{eff}}\omega_m^2\sqrt{S_{x^2}}$ in $\mathrm{J}/\sqrt{\mathrm{Hz}}$. Since $\langle x^2\rangle=x_{\mathrm{zpf}}^2(2n+1)$ for a phonon-number state, the corresponding phonon-number sensitivity is $\sqrt{S_n}=\sqrt{S_{x^2}}/(2x_{\mathrm{zpf}}^2)$ in $\mathrm{phonons}/\sqrt{\mathrm{Hz}}$.

\begin{table*}[t]
\caption{Representative quadratic cavity-transduction platforms, their coupling mechanisms, and comparison with the present split-post resonator.}
\label{tab:quadratic_comparison}
\centering
\begingroup
\setlength{\tabcolsep}{3pt}
\renewcommand{\arraystretch}{1.25}

\begin{tabular}{
@{}
p{0.16\textwidth}
p{0.18\textwidth}
p{0.28\textwidth}
p{0.32\textwidth}
@{}
}
\hline
Platform
& Mechanical element
& Origin of quadratic response
& Distinction of present work \\
\hline

Single-post re-entrant cavity
& Membrane or movable cavity wall
& Primarily linear capacitive frequency shift
& Strong displacement sensitivity, but with first-order linear coupling \\
\hline

Membrane-in-the-middle optical cavity
& Dielectric membrane
& Symmetry point of the optical standing wave
& Optical analogue of symmetry-based quadratic readout \\
\hline

Drumhead NEMS microwave system
& Micro- or nanomechanical drum
& Nonlinear electromechanical coupling and mode geometry
& Very small effective mass and strong coupling, but the geometry is generally fixed after fabrication \\
\hline

Split-post re-entrant cavity
& Dielectric membrane
& Symmetric two-sided re-entrant field cancels first-order coupling while retaining finite curvature
& Macroscopic dielectric membrane, strong microwave-field overlap, and a tunable transition between quadratic and linear response through membrane repositioning \\
\hline
\end{tabular}

\endgroup
\end{table*}

\subsection{Comparison with Other Quadratic Transduction Platforms}

Table~\ref{tab:quadratic_comparison} places the present split-post resonator in the context of representative quadratic cavity optomechanical and microwave transduction platforms. In a conventional single-post re-entrant cavity, displacement changes one dominant capacitive gap and therefore produces a predominantly linear frequency shift. Membrane-in-the-middle optical cavities achieve quadratic coupling by positioning a dielectric membrane at a symmetry point of the optical standing wave, while drumhead NEMS devices exploit their small effective mass and engineered electromechanical mode structure to obtain strong coupling. The split-post resonator implements an analogous symmetry condition in a microwave re-entrant geometry: the first-order frequency shifts produced by the two opposing field regions cancel when the membrane is centred, while a finite second-order curvature remains. Its principal distinction is the combination of strong microwave-field overlap with a macroscopic dielectric membrane and the ability to tune continuously between predominantly quadratic and linear transduction by repositioning the membrane relative to the symmetry point.

\section{\textbf{Conclusion}}

It was demonstrated that the split-post microwave cavity readout exhibited parametric quadratic behaviour due to the intrinsic symmetry of the microwave resonance. When the sapphire membrane was positioned at the centre or off-centre within the microwave resonator, the output response appeared quadratic or linear, respectively. This parametric readout system had the potential to be developed into a transducer capable of measuring energy when operated in its ground state. This work could be scaled to a single-graviton detection scheme, in which kilogram-scale detectors were impedance-matched to gram-scale detectors to measure energy deposited by a gravitational wave at kHz frequencies and thereby probe energy quantisation. Such an approach, when enabled by ground-state cooling of the resonator, could yield promising results, particularly for phonon counting by building strongly coupled detectors capable of measuring energy \cite{aspelmeyer2022zeh, tobar2025detecting}. Our analysis further showed that the split-post resonator had a higher aspect ratio and greater field overlap in the region of the mechanical modes \cite{parashar2024upconversion} than a single-post resonator \cite{parashar2024upconversion,bourhill2020generation}, resulting in a higher output coupling rate. By incorporating these features of split-post resonators, we were able to read out a non-piezoelectric membrane and characterise its response, contributing to the development of a quadratic transducer.

\begin{acknowledgments}
This research was supported by the ARC Centre of Excellence for Dark Matter Particle Physics (CE200100008). Maxim Goryachev is supported by the Australian Research Council Future Fellowship FT250100168.
\end{acknowledgments}

\nocite{*}
\providecommand{\noopsort}[1]{}\providecommand{\singleletter}[1]{#1}%
\providecommand{\newblock}{}

\end{document}